\documentclass[%
reprint,
 amsmath,amssymb,
 aps,pre,longbibliography
]{revtex4-2}

\usepackage{graphicx}
\usepackage{dcolumn}
\usepackage{bm}

\begin{document}

\preprint{APS/123-QED}

\title{Exact amplitude relations for diffusion-limited aggregation}

\author{Thomas C Halsey}

\email{thomas.halsey@rice.edu}
\affiliation{Dept. of Chemical and Biomolecular Engineering, Rice University, Houston, TX}

\date{\today}

\begin{abstract}
It has been known for several decades that the third moment of the multifractal spectrum of the harmonic measure for diffusion-limited aggregates is linked to the underlying fractal dimension of the cluster. We demonstrate, using an argument based on the Hastings-Levitov formulation of diffusion-limited aggregation (DLA) in two dimensions, an even stronger link, connecting the universal amplitude of the third moment to the cluster fractal dimension.  This argument can be used for both the standard circular DLA as well as DLA in a cylinder (i.e., with periodic boundary conditions); in the latter case the relationship of the amplitude to the fractal dimension is weaker, and must be extracted via a scaling analysis.
\end{abstract}

\maketitle


\section{Introduction}

More than four decades after its introduction by Witten and Sander in 1981 \cite{witten1981diffusion}, diffusion-limited aggregation (DLA) remains a largely unsolved puzzle of non-equilibrium statistical physics. Although it was soon recognized that the quite simple Witten-Sander model accounts well for pattern formation in a variety of physical phenomena, including colloidal aggregation, two-phase flow with weak or absent surface tension, and Mn dendrites, to name a few examples, the model has frustrated many techniques that have  proven effective for other non-equilibrium problems.  A variety of statistical physics methods have either been adapted or developed for this problem, including real-space RG methods \cite{pietronero1988theory,erzan1995fixed}, conformal techniques, branching models \cite{halsey1992theory,halsey1994diffusion}, and expansions inspired by traditional $\epsilon$-expansions \cite{hastings2001growth}. In many cases these methods can yield approximate estimates for the DLA fractal dimension and other scaling properties; in no case is there an exact solution, or even a controllable approximation that can practically yield high precision results.

One notable research direction was established by Hastings and Levitov in 1998  \cite{hastings1998laplacian}.  These authors developed a powerful reformulation of DLA in two dimensions, representing the cluster evolution by a clearly defined sequence of iterated conformal maps.  Their methods can also be applied to a wide family of other growth models, notably including the Eden model, a classical model for tumor growth that was first proposed in 1963.  For this latter case, Norris, Turner, E.B. Procaccia, and collaborators have developed a number of rigorous results \cite{norris2012hastings,berger2022growth}.  More recently, E.B. Procaccia and I. Procaccia \cite{procaccia2021dimension}, using the Hastings-Levitov (HL) formulation, demonstrated that the fractal dimension of DLA grown from a ``fiber" is $D=3/2$, as first conjectured by Meakin \cite{meakin1983diffusion}. We should also mention remarkable results of Miller and Sheffield, who defined a family of ``quantum Loewner evolutions", which can be simplistically described as growth models of the Eden or DLA types, but on a fluctuating manifold with a quantum gravity-inspired metric \cite{miller2016quantum}.

There is a paucity of well-established scaling relations for DLA.  The most fruitful direction to develop such relations, at least early in the history of this topic, appeared to be to try to link the fractal dimension $D$ of the cluster to the multifractal spectrum $\tau(q)$ (defined below) of the harmonic measure (for DLA, the growth measure) on the surface of the cluster.  There are three such links that seem to be of particular interest.  The first, due to Makarov, specifies that the ``information dimension" $d \tau /dq \vert_{q=1} = 1$, as a generic property of harmonic measures \cite{makarov1985distortion}. In 1985 Turkevich and Scher proposed that the maximum growth probability on the cluster surface, which should determine the asymptote of $\tau(q \to \infty)$, was linked to the fractal dimension (in two dimensions) via $D =1 + \lim_{q \to \infty} d\tau/dq$ \cite{turkevich1985occupancy}.  Numerical evidence regarding this proposal is inconclusive, and it may be invalidated by the influence of fluctuations.  A third proposal, made by this author, was that $\tau(3) = D$, which has thus far survived comparison with numerical results \cite{halsey1987some,halsey1988scaling}. This latter result is the starting point for this study.

I will first review  the standard DLA model, the Hastings-Levitov reformulation, and the multifractal formalism for the DLA harmonic (or growth) measure. Recalling the argument for $\tau(3) = D$, I will point out a subtle paradox of this argument, in that the universality of a coefficient appearing in this model, with respect to details of the model that should not affect its scaling behavior, is unclear.  This paradox can be resolved through use of the Hastings-Levitov approach, which allows the specification of a universal amplitude, so that both the amplitude and scaling behavior of the third moment of the harmonic measure (in the continuum Hastings-Levitov form) can be directly related to the cluster fractal dimension.  This can be done both for circular DLA, grown from a seed in an unbounded two dimensional domain, as well as for a less studied model ``cylindrical" DLA, in which growth proceeds from a substrate with imposed periodic boundary conditions (unlike the fiber case mentioned above) \cite{evertsz1990self, kol1998solution, benjamini2008diffusion}. For cylindrical DLA the result for the fractal dimension is weaker, since the amplitude does not directly determine this dimension, which must be extracted from a scaling analysis.

I will present numerical results confirming these arguments using the Hastings-Levitov algorithm.  A side note, inspired by these results, is to observe the quite strong fluctuations, seen both in the fractal dimension and in the growth measure moment, for moderate-sized clusters, especially in the circular geometry.

\section{Models and Measures}

The original, and in some ways simplest, description of DLA is as growth from a seed in an unbounded space.  Random walkers arrive from the point at infinity, and aggregate irreversibly when they encounter the cluster. Fractal dimensions for the resulting clusters have been computed in a variety of dimensions $d$; here we focus on $d=2$, for which $D \approx 1.71$.  Lattice effects can be relevant, and can influence this dimension, so we always restrict ourselves to the off-lattice problem.

A growth measure can be defined on the surface of the cluster as the probability $p_i$ that the next particle to aggregate will aggregate to the $i$'th particle in the existing cluster. Up to corrections at the scale of an individual particle, $p_i$ can be interpreted in terms of the solution of Laplace's equation, $\nabla^2 P = 0$, with boundary conditions $P\vert_{\Omega}=0$ on the cluster surface $\Omega$, and constant flux conditions on $P$ at $\infty$. In this interpretation, we write $p_i = \int_{\Omega_i} \hat n \cdot \partial P$, where the integral is over the portion of $\Omega$ corresponding to the $i$'th particle.  Since $p_i$ is a probability, $\sum_{i=1}^n  p_i =1$, where the sum is over all $n$ particles of a particular cluster, and this constraint fixes the flux of $P$ at $\infty$.  The measure defined by $\int \hat n \cdot \partial P$ is a harmonic measure.

The multifractal formalism is applied to DLA through the (discrete) growth measure $\{p_i\}$.  It is observed that sums of the moments of $\{p_i\}$ scale with the cluster size $R$ (which can be taken to be the radius of gyration of the cluster or, preferably, the size of an equivalent electrostatic capacitor, see below), defining an exponent function $\tau(q)$,
\begin{equation}
\sum_{i=1}^n p_i^q= \left ( \frac{a}{R} \right )^{\tau(q)} ,
\end{equation}
with $a$ the radius of an individual particle. If we prefer, we can express this scaling in terms of $n = \left ( \frac{R}{a} \right )^{D}$, where $D$ is the conventionally defined fractal dimension of the cluster,
\begin{equation}
\sum_{i=1}^n p_i^q = \left ( \frac{1}{n} \right )^{\sigma(q)} ,
\end{equation}
with $\sigma(q) = \tau(q) / D$.  Either $\sigma(q)$ or $\tau(q)$ can be converted to a multifractal ``spectrum" $(\alpha, f(\alpha))$ through Legendre transformation, although this will not be necessary here.

Hastings and Levitov (HL) made explicit the conformal structure underlying DLA in $d=2$, thereby opening up a new avenue for studies of this and related problems.  The fundamental object of their formulation is the conformal map $w(z)$ between the exterior of the cluster in space ($w$) and the exterior of the unit circle in the $z$-domain. By the Riemann mapping theorem, the requirement that the cluster surface map to the unit circle is sufficient to determine the full map $w(z)$.

Consider the map corresponding to the $n-1$-particle cluster, $w_{n-1}(z)$. HL specify that the map corresponding to the $n$-particle cluster can be found as
\begin{equation}
w_n(z) = w_{n-1} (f_{\lambda_n,\theta_n}(z)) .
\end{equation}
The function $f$ is somewhat complicated.  First, the parameter $\theta_n$ is a random variable, equidistributed on $[0,2\pi)$.  This corresponds to the choice of the position on the unit circle in the $z$-plane at which the next particle will attach. The parameter $\lambda_n$ scales the size of the object attached through the iteration of the function $f$, ensuring that it is always at the same scale in the $w$-plane through the requirement
\begin{equation}
\lambda_n = \lambda_0 \left \vert \frac{dw_{n-1}}{dz} \right \vert_{z=e^{i\theta_n}}^{-\alpha} ,
\label{eq:lambda_n_def}
\end{equation}
with $\alpha = 2$. Note that this is a lowest-order in $\lambda_0$ result, which can lead to awkwardness deep in the fjords of the cluster, although I shall not concern myself with these complications. Choosing other values of $\alpha$ allows access to other models, such as the Eden model, through the same formalism.

\begin{figure}[t]
\centering
\includegraphics[width=0.45\textwidth]{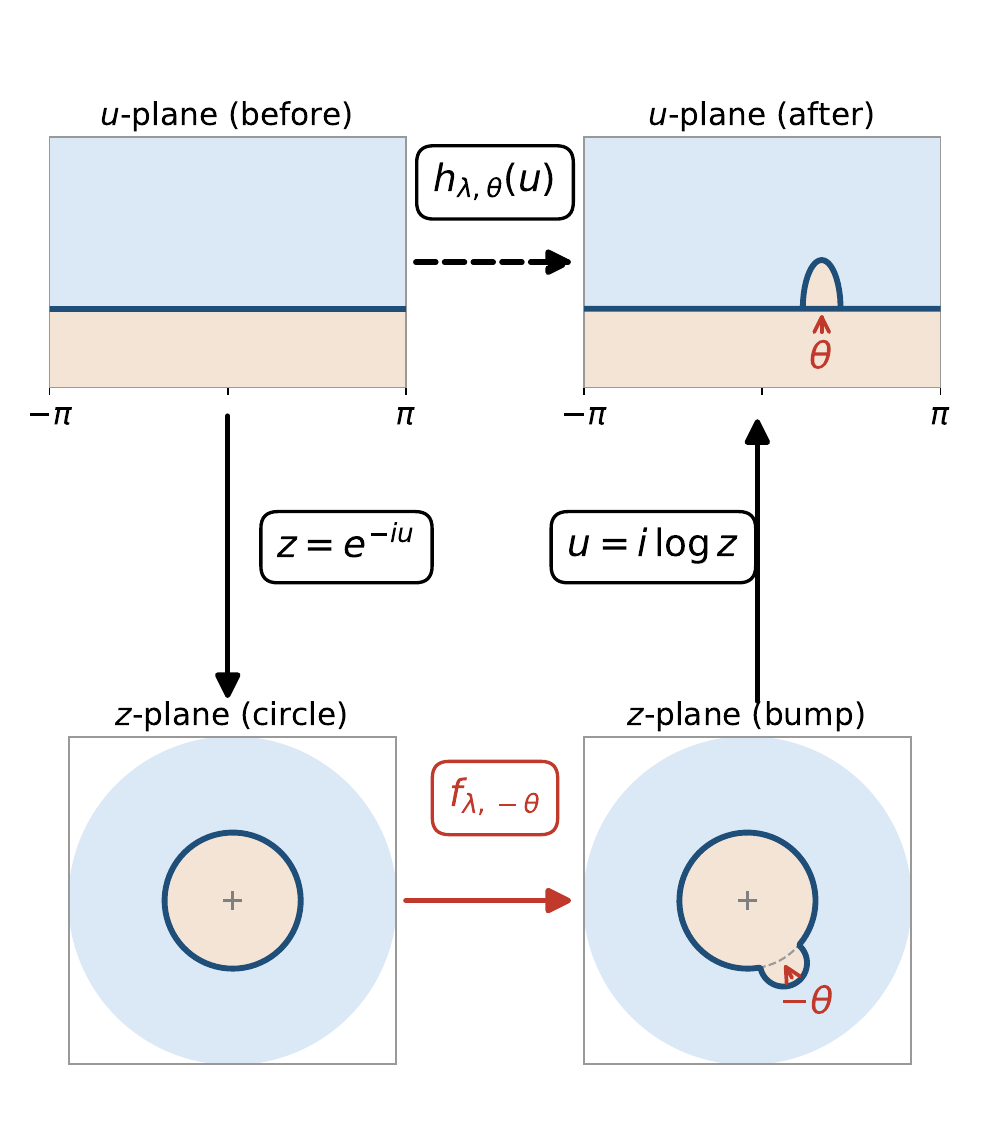}
\caption{\label{fig:cylinder_maps} Nested conformal transformations defining the cylindrical Hastings--Levitov (HL) map $h_{\lambda,\theta}(u) = i\log\!\left( f_{\lambda,-\theta}(e^{-iu}) \right)$. The periodic $u$-plane interface (top left) is sent to the unit circle in the $z$-plane via $z=e^{-iu}$; the planar bump map $f_{\lambda,-\theta}$ attaches a particle at angle $-\theta$; and $u=i\log z$ returns to the cylinder, placing the bump on the interface at ${\rm Re}\,u=\theta$. The dashed arrow is the composite map. The bump is drawn with $a=1/2$, for which the HL particle is a semicircle; the production runs use $a=2/3$. Schematic diagram, rendered from the author's conformal-map formulae.}
\end{figure}

The function $f$ is specified, following the original work of HL, by a function with convenient conformal properties, 
\begin{equation}
f_{\lambda,\theta}(z) = e^{i \theta} f_{{\rm bump},\lambda}(e^{-i\theta}z) ,
\end{equation}
with
\begin{equation}
\begin{split}
f_{{\rm bump},\lambda}(z) ={}& z^{1-a} \bigg[ \frac{1+\lambda}{2z}(z+1) \bigg( z+1 \\
&+\sqrt{z^2+1-2z\,\frac{1-\lambda}{1+\lambda}}\,\bigg) - 1 \bigg]^a ,
\end{split}
\end{equation}
This function creates a bump at the position $z=1$ on the unit circle, with a size parameterized by $\lambda$.  The exponent $a$ is a parameter that determines the ``roundness" of the bump.  To avoid distortions, Ref. \onlinecite{davidovitch1999diffusion} recommends the choice of a value $a = 2/3$, which I use in the numerical studies reported below.

Now the full Hastings-Levitov formulation can be summarized by writing the full iterative conformal map for the $n$-particle cluster
\begin{equation}
w_n(z) = f_{\lambda_1,\theta_1}\left (f_{\lambda_2,\theta_2} \left (f_{\lambda_3,\theta_3}(\cdots f_{\lambda_n,\theta_n}(z)\right ) \right) .
\end{equation}
Note that the most recent particle corresponds to the argument most deeply embedded in the iteration. From the chain rule, it immediately follows that
\begin{equation}
\begin{split}
w^{\prime}(z) ={}& f_{\lambda_1,\theta_1}^{\prime}\left(f_{\lambda_2,\theta_2}(\cdots)\right) \\
&\times f_{\lambda_2,\theta_2}^{\prime}\left(f_{\lambda_3,\theta_3}(\cdots)\right)
\cdots \times f_{\lambda_n,\theta_n}^{\prime}(z) .
\end{split}
\end{equation}

\section{Capacity Scaling}

The $\tau(3) = D$ scaling law, mentioned above, has its roots in the scaling of the capacity of a cluster.  To understand this, it is convenient to replace the flux boundary condition at $\infty$ for the Laplace solution $P$, mentioned above, with a constant potential boundary condition on a circle of radius $R_{\infty} \gg R$.  We must be careful to remember now that the flux of $P$ at the cluster surface can no longer be identified with the growth measure $p_i$, although this latter remains proportional to the flux. In this boundary condition configuration, we can write a general expression for the change in the capacity of the cluster due to infinitesimal changes in its surface position. Reference \onlinecite{halsey1988scaling} gives this formula as
\begin{equation}
\delta \left (\int ds \partial_n P(s) \right ) = \int ds f(s) (\partial_n P)^2 (s)
\label{eq:capacity}
\end{equation}
where the integral is over the cluster surface, $f(s)$ is the normal change in that surface at the location $s$, and $\partial_n P$ is the outward normal derivative of $P$ (the electric field perpendicular to the surface if $P$ is interpreted as an electrical potential).  Now the capacity of a circle of radius $R$ (with respect to another surface at a great distance of radius $R_{\infty}$) is
\begin{equation}
C = \frac{2 \pi}{\log(R_{\infty}/R)}
\end{equation}

If we approximate $f(s) = \bar f a^2 \delta (s - s^{\prime})$ with $s^{\prime}$ the attachment point of the next particle, then we can find the simple formula
\begin{equation}
\frac{d \log R}{d \log n} = 2 \pi n \bar f \sum_i p_i^3 \sim n \left ( \frac{a}{R} \right )^{\tau(3)},
\end{equation}
which immediately gives the scaling law $\tau(3) = D$. Details may be found in References  \onlinecite{halsey1987some,halsey1988scaling}. (Generalization to higher dimensions is straightforward.) However, it is difficult to make further advances here with the normal formulation of DLA, since the particular form of $f(s)$ corresponding to the addition of one particle, and hence the parameter $\bar f$, seems to be highly dependent on microscopic, non-universal details of the model.

The situation is different, however, in the Hastings-Levitov formulation, for which the capacity scaling argument is more powerful. In this formulation, the capacity of the cluster, and hence its characteristic size, can be extracted from the coefficient of the first term in the Laurent expansion for $w(z) = g_1 z + g_0 + g_{-1} z^{-1} + \cdots$, as $g_1 \propto R$, where the scale $R$ is now directly determined by the cluster electrostatic capacity and is thus in a scaling sense independent of other geometrical measures such as the radius of gyration. It immediately follows that for the addition of one particle,
\begin{equation}
\log R_n - \log R_{n-1} = a \log(1+\lambda_n)= a \lambda_n + O(\lambda_n^2)
\end{equation}
which leads to
\begin{equation}
\frac{ \delta \log R}{\delta n} =  a \lambda_0  \left \vert \frac{dw_{n-1}}{dz} \right \vert_{z=e^{i\theta_n}}^{-2}  + O(\lambda_n^2) .
\end{equation}
Averaging over the growth point $\theta$, we obtain at lowest order in $\lambda_n$, or, formally, in $\lambda_0$
\begin{equation}
\left \langle \frac{ \delta \log R}{\delta \log n} \right \rangle = n a \lambda_0 \int \frac {d \theta}{2 \pi}  \left \vert \frac{dw_{n-1}}{dz} \right \vert_{z=e^{i\theta_n}}^{-2}
\end{equation}
and thus the result (again, at lowest order in $\lambda_0$),
\begin{equation}
a \lambda_0 \int \frac {d \theta}{2 \pi} \left \langle \left \vert \frac{dw_{n-1}}{dz} \right \vert^{-2} \right \rangle= \frac{1}{Dn} .
\label{eq:final_circle_result}
\end{equation}
as $n \to \infty$, where we have assumed normal fractal scaling $n \propto R^D$.  Unlike the capacity scaling law of References \onlinecite{halsey1987some,halsey1988scaling}, in this case the amplitude as well as the scaling of the moment integral are constrained.  Equation \ref{eq:final_circle_result}, and its cylindrical companion (below), are the key results of this work.

\section{Cylindrical Geometry}

An interesting variant of two dimensional DLA can be constructed by requiring that the growing cluster be confined to the surface of a cylinder, with the particles introduced at one end of the cylinder far from the growing cluster.  This model has been studied by various authors, and the overall structure found of a fractal, highly dendritic cluster is similar to DLA growth in a plane.  For a fractal dimension $D$, simple scaling suggests that the number of particles in an ultimate cluster of height $H$ will scale with the circumference of the cylinder $L$ as $n \propto  H L^{D-1}$ for $H \gg L$, so that as the cylinder circumference increases, the density of the cluster will decrease (for $D<2$).

We can again use Eq.~(\ref{eq:capacity}) to connect the growth to the third moment of the growth measure. Consider growth in a cylinder of circumference $L$ with an effective gap between the growing cluster and the source of the flux of $Z_{\infty} \gg L$.  The capacitance of the system is $C= L / Z_{\infty}$, which we can in fact use to define $Z_{\infty}$.   Applying Eq.~(\ref{eq:capacity}), we find that analogously to the radial case, with the addition of a particle of size $a$
\begin{equation}
\delta \left ( \frac{Z_{\infty}}{a} \right ) \sim \frac{L}{a} \sum_i p_i^3 = \left ( \frac{a}{L} \right )^{\tau(3)-1} ,
\label{eq:cyl_sum}
\end{equation}
where the sum is over all particles in the cluster.  This is consistent with the claim that the density is $\sim L^{D-2}$ for $D=\tau(3)$, as in the radial case.

An interesting question is whether the dimension $D$ in a cylinder is identical to that found in a plane or not-- numerical works suggest, but do not conclusively demonstrate, that in a cylinder a somewhat lower fractal dimension $D \approx 1.66$ is seen, compared to the planar result $D \approx 1.71$.   This may be linked to suggestions by Mandelbrot and co-authors that in the plane, a dimension $D \approx 1.66$ more accurately characterizes local structure due to ``lacunarity", notwithstanding the well-established result that at large scales $n \propto R^D$ with $D \approx 1.71$. Other authors have questioned this line of argument,  showing evidence that for sufficiently wide channels / cylinders the dimension crosses over to $D \approx 1.71$ \cite{somfai2003channel}.

To assimilate cylindrical growth to the Hastings-Levitov formulation, we specify two planes, with complex variables $u$ and $v$. In both planes, the problem is periodic in the direction of ${\rm  Real}(u)$ or ${ \rm Real}(v)$ with period $2 \pi$.  We now build the conformal map $v(u)$ that describes the cluster surface (for $u = x + i 0$ with $x \in[0,2 \pi)$) by writing
\begin{equation}
v_n(u) = v_{n-1} \left (h_{\lambda_n,\theta} (u) \right ) .
\end{equation}
So far, this is identical to the HL procedure in the plane. To construct the function $h$, we first conformally transform to the plane, then apply the planar HL``bump function" from above, and then transform back to the cylindrical space. (See Fig.~\ref{fig:cylinder_maps}). In particular,
\begin{equation}
h_{\lambda,\theta} (u)  =  i \log \left ( f_{\lambda,-\theta} (\exp( -i u)) \right ) .
\end{equation}
Note that it was necessary to change the sign of $\theta$ to preserve the interpretation that $\theta$ is the real part of the value of $u$ at which attachment of the next particle takes place.

Note that to actually compute conformal maps, we will need the derivative
\begin{equation}
h^{\prime}(u) = \exp(-iu) \frac{f_{\lambda,-\theta}^{\prime} (\exp( -i u)) }{f_{\lambda,-\theta} (\exp( -i u)) } ,
\end{equation}
with
\begin{equation}
\begin{split}
v^{\prime}(u) ={}& h_{\lambda_1,\theta_1}^{\prime}\left(h_{\lambda_2,\theta_2}(\cdots)\right) \\
&\times h_{\lambda_2,\theta_2}^{\prime}\left(h_{\lambda_3,\theta_3}(\cdots)\right)
\cdots \times h_{\lambda_n,\theta_n}^{\prime}(u) .
\end{split}
\end{equation}
This is a somewhat simpler version of an argument originally advanced in Ref.~\onlinecite{somfai2003channel}. 

\begin{figure*}[t]
\centering
\begin{minipage}[b]{0.46\textwidth}
  \centering
  \includegraphics[width=\textwidth]{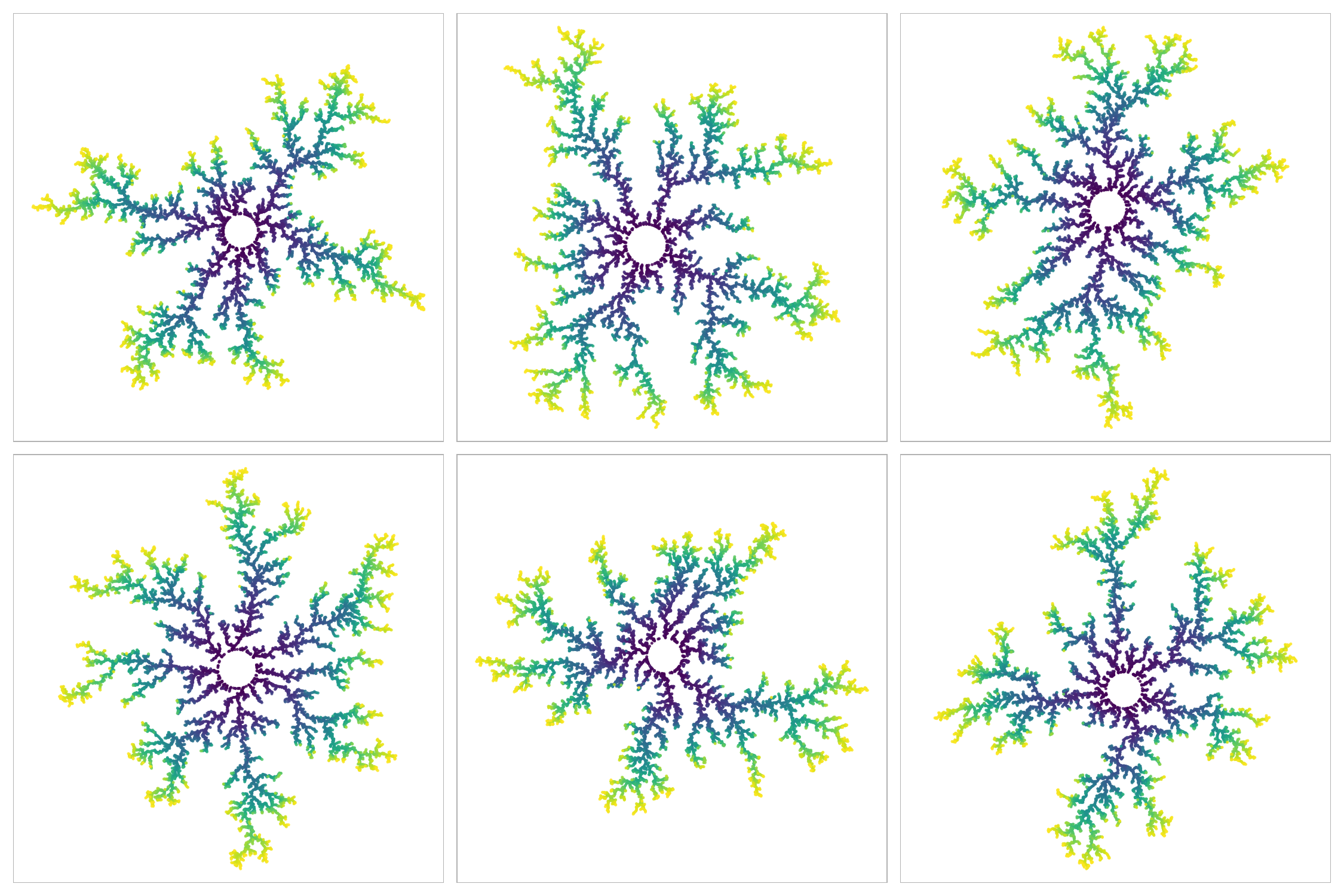}\\[2pt]
  {\small (a)}
\end{minipage}
\hfill
\begin{minipage}[b]{0.50\textwidth}
  \centering
  \includegraphics[width=\textwidth]{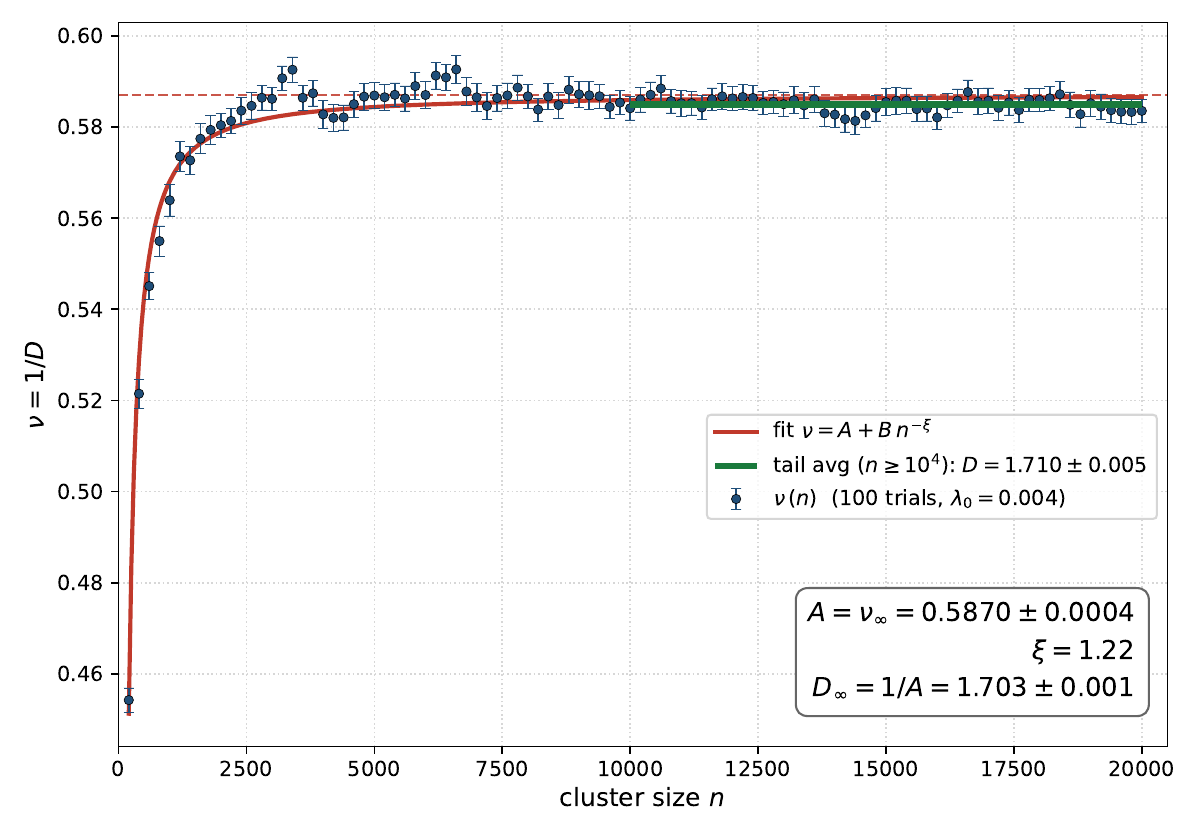}\\[2pt]
  {\small (b)}
\end{minipage}
\caption{\label{fig:circle} (a) Six randomly chosen DLA clusters generated by the
Hastings--Levitov algorithm with $\lambda_0=10^{-3}$, $a=2/3$, and $n=10{,}000$;
the color indexes the order of attachment, with considerable variation in gross
shape. (b) Computed $\nu(n)$ from Eq.~(\ref{eq:dimension_fit}); averaging over
$10{,}000<n<20{,}000$ gives $D=1.71$, while a fit to the full curve gives a slightly
lower $D(n\to\infty)\equiv D_\infty$.}
\end{figure*}

Equivalently to Eq.~(\ref{eq:lambda_n_def}), we have
\begin{equation}
\lambda_n = \lambda_0 \left \vert \frac{dv_{n-1}}{du} \right \vert_{u=\theta_n}^{-2} ,
\end{equation}

It is now direct to see the change in $v(u)$ for ${\rm Im}(u)$ large corresponding to the addition of one particle.  The general form of $v(u)$ for $u \to +i \infty$ is
\begin{equation}
v(u) = u + v^{(0)} + v^{(1)} \exp(i u) + \cdots .
\end{equation}
Direct substitution then yields
\begin{equation}
v_{n}^{(0)} = v_{n-1}^{(0)} + i a \lambda_n +O(\lambda_n^2) .
\end{equation}
Thus the equation of motion of the effective height $H$ of the interface (defined by the value of $v_{n}^{(0)}$), is
\begin{equation}
\frac{dH}{dn} = { \rm Im}  [ \frac{d v_{n}^{(0)}}{dn} ] = a \lambda_0 \int \frac {d \theta}{2 \pi} \left \langle \left \vert \frac{dv_{n-1}}{du} \right \vert_{u=\theta}^{-2} \right \rangle ,
\label{eq:cyl_int}
\end{equation}
Although this is the exact $O(\lambda_0)$ term for the effective velocity of the growing interface, extraction of a fractal dimension requires a scaling analysis, unlike the case for radial growth.  This analysis can be based on the formula
\begin{equation}
\frac{d H}{dn} \propto \lambda_0^{D/2} ,
\label{eq:cyl_scaling}
\end{equation}
by using computations with various values of $\lambda_0$.

\section{Numerical Results}

\subsection{Circular Geometry}

\begin{figure*}[t]
\centering
\begin{minipage}[c]{0.46\textwidth}
\centering
\includegraphics[width=\textwidth]{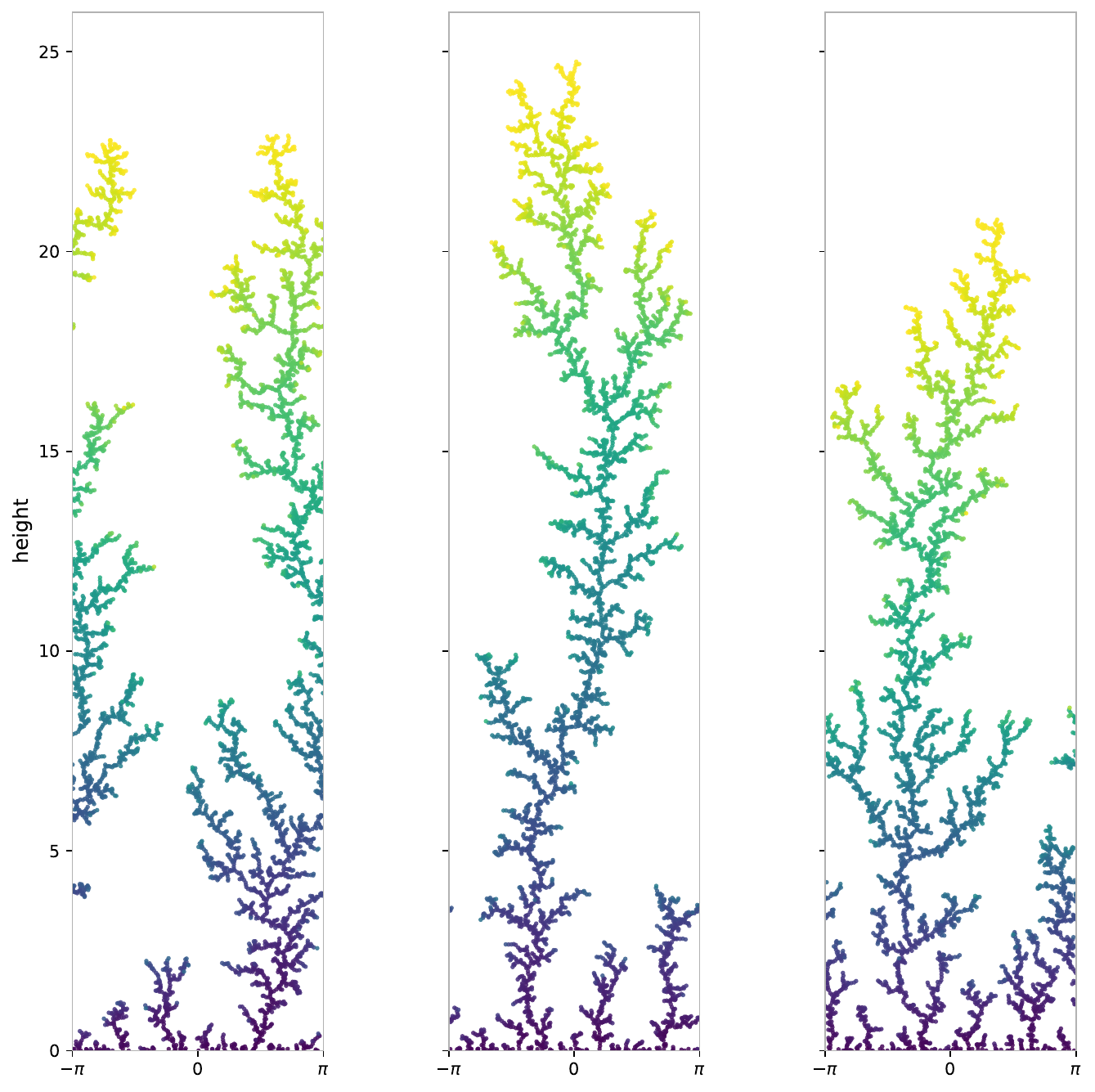}\\[2pt]
(a)
\end{minipage}
\hfill
\begin{minipage}[c]{0.50\textwidth}
\centering
\includegraphics[width=\textwidth]{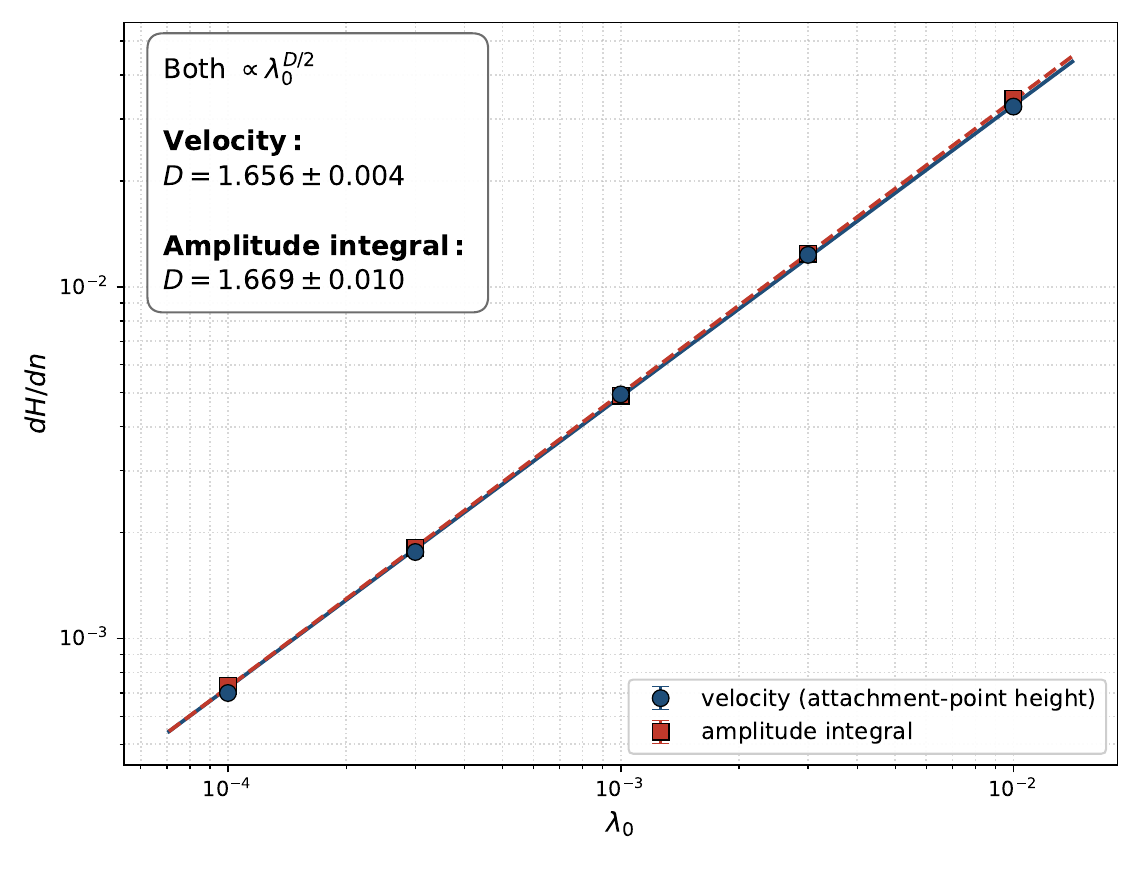}\\[2pt]
(b)
\end{minipage}
\caption{\label{fig:cylinder} Cylindrical DLA. (a) Three randomly chosen clusters generated by the Hastings-Levitov algorithm with $\lambda_0 = 10^{-3}$, $a=2/3$, and $n=5,000$; the color indexes the order of attachment of the particles, and there is less variation in the gross shape of the clusters than in circular DLA. (b) The scaling of the harmonic measure sum from Eq.~(\ref{eq:cyl_int}) determines the overall fractal dimension of the cluster, via Eq.~(\ref{eq:cyl_scaling}); the result $D=1.67$ is consistent with the literature. We also show a more direct fractal dimension determination from the slope of the height of the attachment points versus $n$.}
\end{figure*}

The use of the Hastings-Levitov method to develop DLA clusters in a circular geometry has been discussed, in detail, by a number of authors.  Following Ref.~\cite{davidovitch1999diffusion}, we choose $a = 2/3$ and vary $\lambda_0$ (effectively, particle size) as well as the number of particles, $n$, in a cluster.  The overall algorithm is $O(n^2)$, which means that serious computational resources are required for large clusters (the computations on which this section is based were performed on a MacBook M1, which was adequate for computing the amplitudes discussed above for moderately-sized clusters).  Figure~\ref{fig:circle}(a) shows results (chosen at random) for six clusters of $n=10{,}000$ particles for $\lambda_0=10^{-3}$.   The original circular ``seed" from which the clusters were grown is clearly visible at the center.

An advantage of the HL approach is that the harmonic measure is computed as part of the process of instantiating the cluster, so it is direct to compute the dimension from Eq. (\ref{eq:final_circle_result}).  This equation has a very simple dependence on $n$, which allows the dimension to be essentially read off of numerical results.  One practical complication in doing this is that Eq. (\ref{eq:final_circle_result}) is true only to lowest order in $\lambda_0$, so one would prefer to test this prediction at small values of $\lambda_0$.  Since $\sqrt{\lambda_0}$ is essentially the size of a particle, these smaller values require many more particles to escape the influence of the original circle from which the growth starts.  Our solution is to choose a moderate value of $\lambda_0 = 0.004$ and grow 100 instances out to $n=20{,}000$ particles, averaging the left-hand side of  Eq. (\ref{eq:final_circle_result}), and then fitting to a constant plus a power law,
\begin{equation}
a \lambda_0 n \int \frac {d \theta}{2 \pi} \left \langle \left \vert \frac{dw_{n-1}}{dz} \right \vert^{-2} \right \rangle \equiv \nu(n) = A+ Bn^{-\xi} ,
\label{eq:dimension_fit}
\end{equation}
reading the ultimate dimension as the inverse of the plateau value $A$.   Results are shown in Fig.~\ref{fig:circle}(b); the result is $D(n \to \infty) \equiv D_{\infty} =1.703 \pm 0.001$.  Note that this is statistical error only, and neglects structural error arising from inadequacies of the fitting form.  Alternatively, one can directly average the results for the plateau value $10{,}000 < n < 20{,}000$,  yielding $D= 1.710 \pm 0.005$, which is within one standard deviation of  the accepted value $D=1.713 \pm 0.003$ \cite{davidovitch2000convergent}.  Note the strong fluctuations seen in these results.

\subsection{Cylindrical Geometry}

Now we turn to DLA in a cylindrical geometry.  Despite the more complicated construction of the conformal map in the HL method, this is in some ways a simpler problem, because the cluster rapidly asymptotes to its apparent large $n$  structure, in practice as soon as its aspect ratio significantly exceeds one (see Fig.~\ref{fig:cylinder}(a)).  This simplifies the computation of the velocity $\langle d H/ dn \rangle \propto \langle d v^{(0)}/dn \rangle$ in either a direct or a harmonic measure-based method.  The direct method averages the slope of the attachment heights of particles vs. $n$.  Results are shown in Fig.~\ref{fig:cylinder}(b), and are consistent with the literature values of $D \approx 1.66$ \cite{kol1998solution,taloni2006conformal}.  Note, however that Ref.~\onlinecite{somfai2003channel} claims that for very large cylindrical clusters, the dimension crosses over to $D \approx 1.71$.  This occurs for cylinder (or channel) widths $\sim 10^3$ particle diameters, corresponding to $\lambda_0 \sim 10^{-6}$, well below the values I have studied here. 

\hskip .1 in
\begin{acknowledgments}
I am grateful to Bertrand Duplantier and Matthew Von Hippel for an invitation to visit Institut de Physique Th\' eorique, Universit\' e de Paris, Saclay, at which this work was performed.  The author used Claude (Cowork; Opus 4.8; Anthropic) in preparing this manuscript: to optimize the simulation code; to generate and format the figures from Hastings–Levitov data produced by the author; and to assist with LaTeX/BibTeX troubleshooting, figure placement, and equation formatting. This includes the schematic in Fig.~1, which was rendered by author-directed code from conformal-map formulae derived by the author. All such work was carried out under the author's direction; the figures present the author's own simulation data and formulae and contain no AI-generated photographic or data imagery. The tool contributed no independent scientific content; all output was reviewed and verified by the author, who takes full responsibility for the work.

\end{acknowledgments}

\bibliography{dla_new}

\end{document}